\documentclass[a4paper]{jpconf}
\usepackage{graphicx}
\begin{document}
\title{Ludendorff Coronal Flattening Index of the Total Solar Eclipse on March 9, 2016}

\author{Tiar Dani\textsuperscript{a*)}, Rhorom Priyatikanto\textsuperscript{a)}, and Abdul Rachman\textsuperscript{a}}

\address{\textsuperscript{a)}Space Science Center, National Institute of Aeronautics and Space, LAPAN
	Jl. Dr. Djundjunan 133 Bandung 40173, Indonesia
	}

\ead{*tiar.dani@lapan.go.id}

\begin{abstract}
Ludendorff coronal flattening index of the Total Solar Eclipse (TSE) on March 9, 2016, was calculated at various distances in solar radius. As a result, we obtained the coronal flattening index $\left(\epsilon =a+b\right)$ at a distance of 2 solar radii is 0.16. The $24^{th}$ solar cycle phase based on the 2016 TSE event obtained -0.64 which showed the corona is pre-minimum type. Resulted coronal flattening index value gives a predicted maximum amplitude of the monthly sunspot number ($W_{max}$) for the $25^{th}$ solar cycle to be $70\pm65$. Therefore, the solar activity for $25^{th}$ solar cycle predicted to be lower than the current solar cycle, which has a maximum sunspot number value of 146 in February 2014
\end{abstract}

\section{Introduction}
Space-based observations provide opportunity to study the physics of solar corona \cite{Brueck1995}, but ground-based observations conducted during total solar eclipses give another chances to understand the lower part of corona  \cite{Pas2007} since the global and local magnetic field of the sun inner part influenced the solar coronal shapes. Consequently, the structure will change according to the solar cycle. Corona at minimum solar activity has asymmetrical brightness distribution characteristics around the solar disc with two very bright streamers around the equator. While corona at maximum solar activity has characteristics of uniform brightness with the streamer around the sun \cite{Sad2007}. Photometry of the K and F corona observed using the white-light filter during TSE events \cite{Sto2007}. K-corona is the photosphere light scattered by the free electron gas in the corona while F-corona is the photosphere light scattered by dust particles of small size in the ecliptic plane \cite{ngging}.

One way to quantify the global structure of the corona is using Ludendorff coronal flattening index ($\epsilon $) \cite{Sto2011}. Several studies calculated coronal flattening index for TSE event have been conducted \cite{Gul1997,Sto2007,Sad2007,Pish2011,Sto2011,Sto2012}. From the compilation of measurements, variation of coronal flattening index over solar cycle can be modeled and the characteristics of upcoming solar cycle (e.g. time of maximum and the amplitude) can be predicted \cite{Pish2011}. In this study, we obtained coronal flattening index by analyzed the solar image isophotes during totality on the 2016 TSE. Henceforth, the sunspot number maximum amplitude for the next solar cycle is predicted using the coronal flattening index. In addition, the type of corona determined based on the current solar cycle phase.

\section{\label{met}Data and Method}
We conducted the 2016 TSE observation at Ngata Baru ($0^{o}55'27"$ South, $119^{o}57'27"$ East, $320m$ asl.), Central Sulawesi, Indonesia. At this point, the duration of totality is 2m 14s while the general sky condition was clear. Four images were acquired using 12 Mpx camera with various field of views and exposure times. All these images were captured using ISO 800 setting with exposure from $1/60s, 1/100s, 1/160s, and 1/200s$. Taken from 07:39:17 until 07:40:41 local time.

No special data reduction was applied, but the images were rotated to get the proper orientation and matched with reference images (SDO/AIA 193 from http://sdo.gsfc.nasa.gov and SOHO/LASCO C2 from http://sohodata/nascom.nasa.gov). Observed features such as prominence and streamer became visual guides for the matching process. Additional data of monthly sunspot number from WDC-SILSO from Royal Observatory of Belgium (http://www.sidc.be/silso) and multi-year forecast sunspot number from SWPC NOAA (http://www.swpc.noaa.gov) was used to determine the phase of current solar cycle.

Ludendorff flattening index was determined from the isophotal contours (isophote in brief) with various equatorial distances from the limb. The isophote was constructed for each grayscale images that was properly oriented to the solar equator. Image with different exposures enable us to construct several sets of isophote that give the value of flattening indexes at various distances. The flattening index itself is defined as

\begin{equation}
\epsilon =\frac{{d}_{0}+{d}_{1}+{d}_{2}}{{D}_{0}+{{D}_{1}+{D}_{2}}}{-1}
\label{eq:flattening}
\end{equation}

\noindent where $d_{0}$ is the diameter of isophote at the equator, $d_{1}$ and $d_{2}$ is the diameter at an angle of $\pm 22.5^{\circ}$ axis parallel to the equator. While $D_{0}$, $D_{1}$, and $D_{2}$ is the diameter isophote in the polar axis direction. The value of $\epsilon$ increase linearly from the limb to a certain distance, $R_{max}=1.5 - 2.5$ $R_{sun}$ then decrease up to a distance $R_{min}=3.5 - 5.5$ $R_{sun}$. Ludendorff (and the citing authors) use $\epsilon$ at $r = 2$ $R_{sun}$ as the photometric flattening index. To obtain this number, extrapolation is often used since the low intensity of the outer part of the solar corona.

The value of $\epsilon $ is also a function of the solar cycle phase ($\Phi$) as introduced by \cite{Lud1928} with following equation:

\begin{equation}
\Phi =\frac{{T}_{ecl}-{T}_{min}}{\left|{T}_{max}-{T}_{min}\right|}
\label{eq:phase}
\end{equation}

\noindent where $T_{ecl}$ is the time of the eclipse (in years), $T_{max}$ and $T_{min}$ is the time of maximum and minimum solar cycle near $T_{ecl}$, respectively. $T_{max}$ identified from WDC-SILSO data and $T_{min}$ identified from SWPC NOAA. Phase of solar activity $\phi$ changes from -1 to 1 and can be interpreted as the phase of solar cycle from maximum to maximum. \cite{Pish2011} obtained a linear fitting equation for the maximum monthly sunspot number ($W_{max}$) with flattening index ($\epsilon$) when approaching the minimum solar cycle using data from 60 events of TSE. This prediction equation has large $1\sigma$-uncertainty ($\pm65$) due mainly to the large sensitivity of the flattening index value to many factors, such as number of innermost isophotes selected for linear approximation, possible error in the orientation of coronal images, etc.(\cite{Pish2011})

\begin{equation}
{{W}_{max}=-2.8 + 466.1 {\epsilon}}
\label{eq:wmax}
\end{equation}

By using Equation \ref{eq:wmax}, the maximum amplitude of the monthly sunspot numbers for the next solar cycle can be calculated. We also compared our results of the coronal flattening index, TSE phase, and $W_{max}$ with other studies.

\section{Results and Discussion} 
\begin{figure}[h]
	\begin{minipage}{38pc}
		\centering
		\includegraphics[width=25pc]{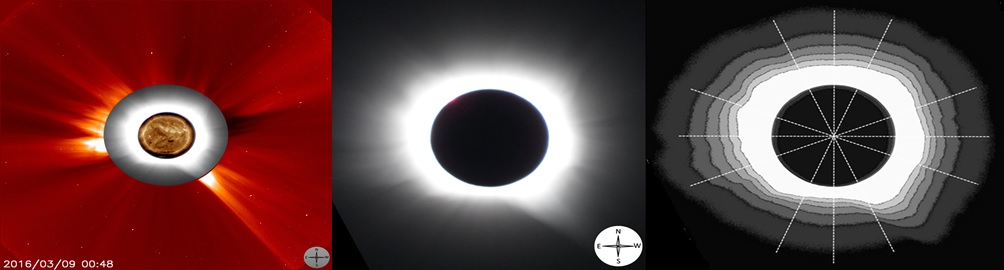}
		\caption{\label{flat}Left panel: TSE image is superimposed on the relevant X-ray image from SDO/AIA193 (inner) and streamer structure from SOHO/LASCO to confirm the exact rotation. Middle panel: Example of the white-light corona during 2016 TSE from our observation with f/5.6 aperture and 1/60s exposure. Right panel: isophotes images with marking line for measurement of flattening index .}
	\end{minipage}\hspace{2pc}%
\end{figure}

The north-south direction of the four TSE 2016 observation images were rotated by $66^{\circ}$ clockwise (\Fref{flat}: Left panel) to get the proper orientation and matched with reference images (see Section \ref{met})

\begin{figure}[h]
	\centering
	\begin{minipage}[b]{18pc}
		\centering
		\includegraphics[width=17pc]{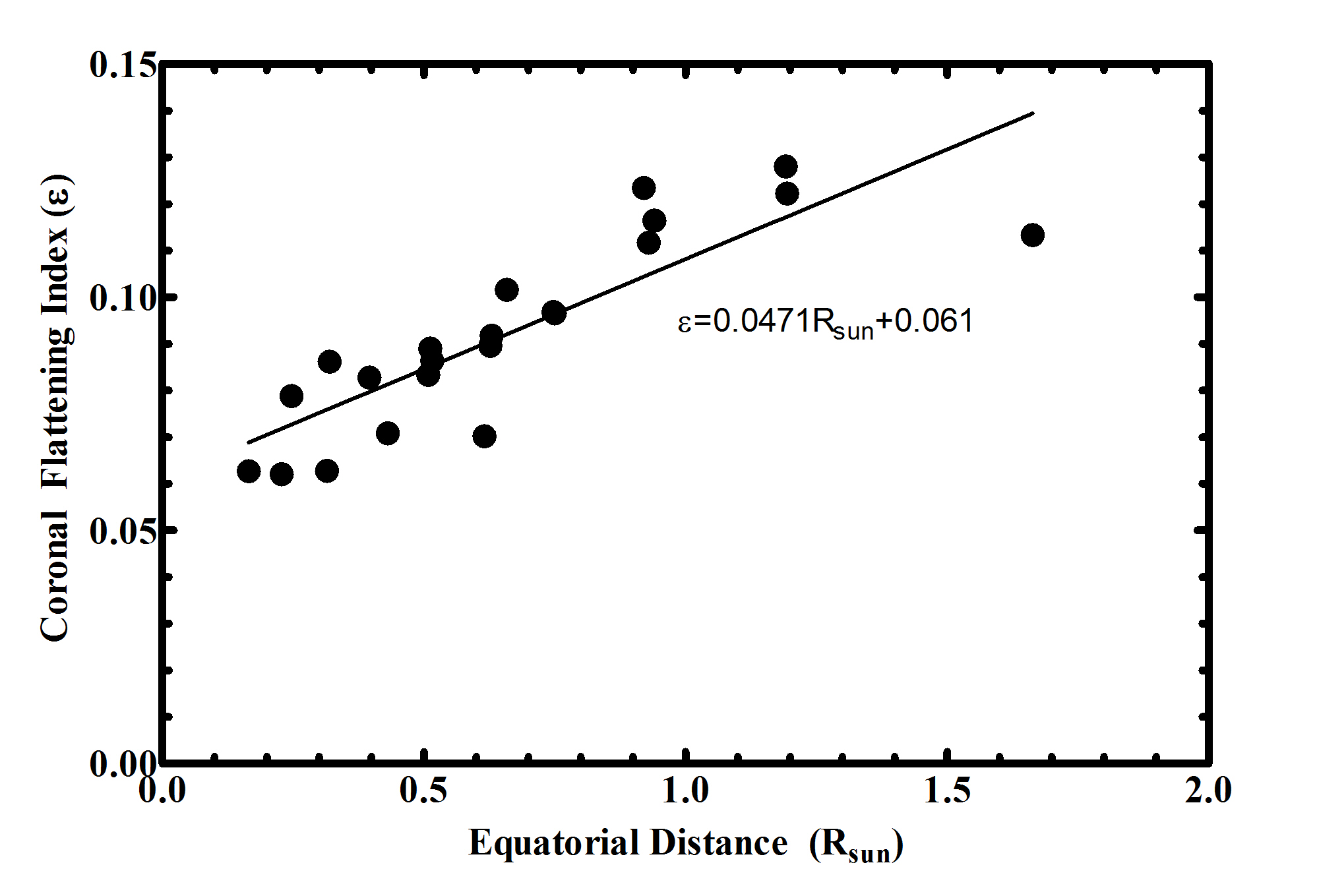}\hspace{2pc}
		\caption{\label{flatplot}Coronal Flattening Index of the solar corona during 2016 TSE.}
	\end{minipage}\hspace{1pc}%
	\begin{minipage}[b]{15pc}
		\centering
		\includegraphics[width=15pc]{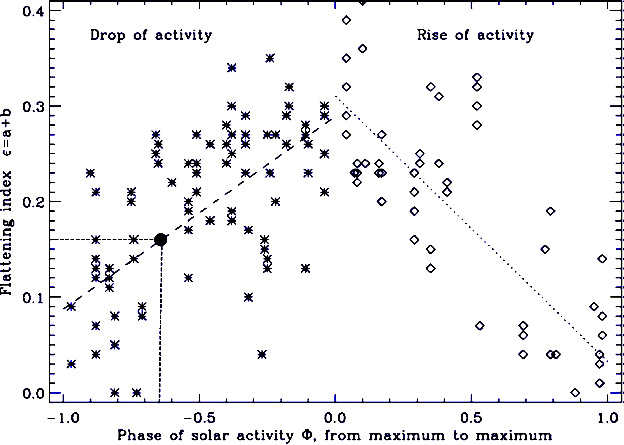}
		\caption{\label{Pish}Dependence of the flattening index on the phase of solar activity \cite{Pish2011}. Result of TSE 2016 shown in black dot mark.}
	\end{minipage}	
\end{figure}

\noindent Twenty isophotes in various solar radius obtained from these four images. Then each of the isophotes calculates its coronal flattening index using Equation \ref{eq:flattening} (\Fref{flat}: Right panel). Coronal flattening index for various solar radius from 20 isophote are plotted and shown in \Fref{flatplot}. The isophotes construction is limited to the distance of ~1.7$R_{sun}$ due to low signal to noise ratio outside this distance. To overcome this limitation, Ludendorff coronal flattening index at $r=2R_{sun}$ is extrapolated using the linear equation $\epsilon=0.0471{R}_{sun}+{0.061}$ obtained from linear regression as displayed in \Fref{flatplot}. Ludendorff coronal flattening index for 2016 TSE is 0.16. 

The 2016 TSE corona phase for $24^{th}$ solar cycle then calculated using Equation \ref{eq:phase}. The peak time (in year) of the $24^{th}$ solar cycle ($T_{max}$) based on data from WDC-SILSO Royal Observatory of Belgium, Brussels, the maximum monthly sunspot number for the $24^{th}$ solar cycle is 146.1 which occurred in February 2014 ($T_{max} = 2014.123$). While the $24^{th}$ solar cycle minimum based on multi-year predictions of monthly sunspot numbers from NOAA SWPC occurred in December 2019 with the monthly sunspot number values is 4.1, so $T_{min} = 2019.923$. And for the eclipse time ($T_{ecl}$) occurred on March 2016 is 2016.206. $24^{th}$ solar cycle phase then calculated using Equation \ref{eq:phase}, obtained $\Phi  = -0.64$. Keep in mind that the solar cycle phase $\Phi$ value is determined by the prediction of $T_{min}$ which might be still get corrected. The value of $\epsilon$ and $\phi$ for TSE in 2016 were added to the plot of coronal flattening index and the solar cycle phase which compile based on the 60 TSE events from 1851 to 2010 as presented in the \Fref{Pish}.

\Fref{Pish} shows that 2016 TSE occurs when solar activity went to its minimum value hence the observed corona is a pre-minimum type corona. Also the 2016 TSE corona shapes are nearly symmetric based on the flattening index value despite a fairly large streamer near the solar south pole on 2016 TSE which affects the solar poles diameter but in this case, does not really influence the final value of the coronal flattening index. 

Furthermore, the prediction for maximum amplitude of the sunspot numbers for $25^{th}$ solar cycle obtained by using Equation \ref{eq:wmax}. So the predicted of maximum sunspot number for the $25^{th}$ solar cycle ($W_{max}$) is $70\pm65$. The flattening index factors used to predict the maximum amplitude sunspot numbers for the next cycle can be regarded as an indirect characteristic of the solar poles magnetic field \cite{Pish2011}. Our result means that the $25^{th}$ solar cycle activity is predicted to be lower than the current cycle, which has peaked in February 2014 with a sunspot number value of 146. This can be compared with results from other studies. For example, Li et al. \cite{Li2015} stated that the next solar cycle will reach maximum on October 2023 with amplitude of 109 which is inside $1\sigma$-uncertainty of our calculated amplitude. Rigozo et al. \cite{Rig2011} also gave slightly higher predicted maximum amplitude of sunspot number in cycle 25 which is 132. This maximum value will be reached on April 2023. Javaraiah \cite{Jav2015} used long-term record of the sunspot groups area to study the solar cycle behavior. The implication of that study is the predicted maximum amplitude of $25^{th}$ solar cycle which is as low as $50\pm10$.

\section{Conclusion}
The coronal flattening index is 0.16 and the $24^{th}$ solar cycle phase is -0.64 based on 2016 TSE observations. This results suggest that the corona is a pre-minimum type. This gives the predicted maximum of the monthly sunspot number for the $25^{th}$ solar cycle to be $70\pm65$. Therefore, the solar activity for $25^{th}$ solar cycle is predicted to be lower than current solar cycle.

\ack
The authors wish to express their thanks to WDC-SILSO, Royal Observatory of Belgium, Brussels, US SWPC NOAA, NASA/SDO and the AIA, EVE, and HMI science teams, NASA and ESA SOHO for providing the solar data.

\section*{References}
\bibliography{ISSEL}

\providecommand{\newblock}{}
\begin{thebibliography}{10}
\expandafter\ifx\csname url\endcsname\relax
  \def\url#1{{\tt #1}}\fi
\expandafter\ifx\csname urlprefix\endcsname\relax\def\urlprefix{URL }\fi
\providecommand{\eprint}[2][]{\url{#2}}

\bibitem{Brueck1995}
Brueckner G~E, Howard R~A, Koomenand M~J, Korendyke C~M, Michels D~J, Moses
  J~D, Socker D~G, Dere K~P, Lamy P~L, Llebaria A, Bout M~V, Schwenn R, Simnett
  G~M, Bedford D~K and Eyles C~J 1995 {\em Solar Physics\/} {\bf 162} 356--402

\bibitem{Pas2007}
Pasachoff J~M, Rusin V, Drucmuller M and Saniga M 2007 {\em The Astrophysical
  Journal\/} {\bf 665} 824--829

\bibitem{Sad2007}
Sadovenko I~V and Pishkalo M~I 2007 {\em Proc. of Contributed Papers WDS'07\/}
  III pp 32--35

\bibitem{Sto2007}
Stoeva P, Stoev A, Kuzin S, Shopov Y, Kiskinova N, N N~S and Pertsov A 2007
  {\em Atmospheric and Solar Terestrial Physics\/} {\bf 70} 414--419

\bibitem{ngging}
Sungging E~M 2015 {\em Media Dirgantara\/} {\bf 10} 7--10

\bibitem{Sto2011}
Stoeva P~V, Stoev A~D and Kuzin S~V 2011 {\em Sun and Geopshere\/} {\bf 6}
  36--38

\bibitem{Gul1997}
Gulyaev R~A 1997 {\em Astrophysical Astronomical and Transactions\/} {\bf 13}
  137--144

\bibitem{Pish2011}
Pishkalo M 2011 {\em Solar Physics\/} {\bf 270} 347

\bibitem{Sto2012}
Stoeva P, Stoev A and Kuzin S 2012 {\em Sun and Geosphere\/} {\bf 7} 81--84

\bibitem{Lud1928}
Ludendorff H 1928 {\em Sitzungsber. Preuss. Akad. Wiss. Phys.-Math. Kl.\/} {\bf
  16} 185

\bibitem{Li2015}
Li K~J, Feng W and Li F~Y 2015 {\em Journal of Atmospheric and
  Solar-Terrestrial Physics\/} {\bf 135} 72--76

\bibitem{Rig2011}
Rigozo N~R, {Souza Echer} M~P, Evangelista H, Nordemann D~J~R and Echer E 2011
  {\em Journal of Atmospheric and Solar-Terrestrial Physics\/} {\bf 73}
  1294--1299

\bibitem{Jav2015}
Javaraiah J 2015 {\em New Astronomy\/} {\bf 34} 54--64

\end{thebibliography}

\end{document}